\documentclass[12pt,preprint]{aastex}
\usepackage{emulateapj5}

\slugcomment{Accepted by ApJ Letters}

\shortauthors{A. K. Inoue}
\shorttitle{SFR from IR}

\begin{document}

\title{Star Formation Rate from Dust Infrared Emission}

\author{Akio K. INOUE}
\affil{Department of Astronomy, Faculty of Science, Kyoto University,
Sakyo-ku, Kyoto 606-8502, JAPAN}
\email{inoue@kusastro.kyoto-u.ac.jp}

\begin{abstract}
We examine what types of galaxies the conversion formula from dust
 infrared (IR) luminosity into the star formation rate (SFR) derived by
 Kennicutt (1998) is applicable to.
The ratio of the observed IR luminosity, $L_{\rm IR}$, to the intrinsic
 bolometric luminosity of the newly ($\la$ 10 Myr) formed stars, 
 $L_{\rm SF}$, of a galaxy can be determined by a mean dust opacity in
 the interstellar medium and the activity of the current star formation.
We find that these parameters area being $0.5 \le L_{\rm IR}/L_{\rm SF} 
 \le 2.0$ is very large, and many nearby normal and active star-forming
 galaxies really fall in this area.
It results from offsetting two effects of a small dust opacity and a
 large cirrus contribution of normal galaxies relative to starburst
 galaxies on the conversion of the stellar emission into the dust IR
 emission.
In conclusion, the SFR determined from the IR luminosity under the
 assumption of $L_{\rm IR}=L_{\rm SF}$ like Kennicutt (1998) is reliable
 within a factor of 2 for all galaxies except for dust rich but quiescent
 galaxies and extremely dust poor galaxies.
\end{abstract}

\keywords{dust, extinction  --- galaxies: general --- infrared: galaxies
--- stars: formation}

\section{Introduction}

Dust infrared (IR) luminosity\footnote{In this Letter, the term 'IR'
means the whole wavelength range of dust emission. The 'IR luminosity'
dose not include the component emitted by stars directly.} is a
familiar indicator of the current star formation rate (SFR).
Especially, the most active star-forming galaxies (e.g., Arp 220) emit
almost all their radiation energy in the IR range \citep{soi87,san96}.
For such galaxies, \cite{ken98a} derived a conversion formula from the
IR luminosity to the SFR by assuming the IR luminosity equal to the
bolometric luminosity and 10--100 Myr continuous star formation
(hereafter K98 IR formula).
More recently, Inoue, Hirashita, \& Kamaya (2000; hereafter IHK00)
theoretically constructed a new conversion law from IR to SFR, which we
can apply not only to dusty starbursts but also to normal (or quiescent)
galaxies.

Although the SFRs determined from various indicators should agree with
each other, we often find discrepancies among them.
Thus, many authors try to resolve such discrepancies
\citep{sul00,bel01,hop01,cha02,bua02,ros02}.
In this context, the SFR from the IR luminosity via K98 IR formula
is assumed to be the correct SFR in spite of the fact that K98 IR
formula can be applied to only dusty starbursts in principle
\citep{hop01,ros02}.
Indeed, the SFR via K98 IR formula shows a good agreement with that
from the H$\alpha$ luminosity even for samples of normal galaxies if we
correct the H$\alpha$ luminosity for the dust extinction in the
interstellar medium (ISM), that in H {\sc ii} regions, and the stellar
Balmer absorption \citep{cha02,ros02}.
Also, an empirical conversion factor into the SFR from the IR luminosity 
for normal late-type spiral galaxies agrees with that of K98 IR
formula within a factor of 2 \citep{bua96}.
Why can K98 IR formula present the correct SFR even for normal galaxies?

In this Letter, we examine what types of galaxies K98 IR formula is
applicable to, and how much correction factor against the formula is
required to obtain a more precise SFR.
In \S 2, we formulate the relation between the IR luminosity and the
star formation (SF) luminosity.
In \S 3, we describe a model of dust extinction of galaxies.
In \S 4, we show that the ratio of the IR luminosity to the SF
luminosity becomes of order unity within a factor of 2 for a very wide
parameter area, and many nearby galaxies really fall in this area.

\section{Infrared Luminosity of Galaxies}

Following IHK00, we formulate the relation between the observed IR
luminosity of a galaxy, $L_{\rm IR}$, and the intrinsic bolometric
luminosity of the newly ($\la$ 10 Myr) formed stellar population
(hereafter SF population), $L_{\rm SF}$.

The IR luminosity of a galaxy is expressed by the following:
\begin{equation}
 L_{\rm IR}=L_{\rm Ly \alpha}+(1-f)L_{\rm LC}+\epsilon L_{\rm UV}
            +\eta L_{\rm under}\,,
 \label{eq1}
\end{equation}
where $L_{\rm Ly \alpha}$, $L_{\rm LC}$, $L_{\rm UV}$, and $L_{\rm
under}$ are luminosities of Lyman $\alpha$ photons generated in H {\sc
ii} regions, Lyman continuum (LC) photons and nonionizing ($\lambda >
912$ \AA) photons emitted by the SF population, and photons from
underlying stellar population, respectively.
The parameters, $f$, $\epsilon$, and $\eta$ are a fraction of LC
photons contributing to hydrogen ionization, an efficiency of the
dust extinction for nonionizing photons from the SF population, and that
for photons from underlying population.
Here, we assume that all Lyman $\alpha$ photons are eventually absorbed
by dust during a number of resonance scatterings in the ISM.
We neglect the escape of LC photons from a galaxy, the consumption of LC
photons by Helium ionization, any extinctions for hydrogen's
recombination lines within H {\sc ii} regions except for Lyman
$\alpha$, and the contribution to the IR luminosity by AGN activity.

According to the Kurucz ATLAS 9 stellar spectra with Solar metallicity
and a turbulence speed of 2 km s$^{-1}$, we obtain $L_{\rm
LC}=0.37L_{\rm SF}$ and $L_{\rm UV}=0.63L_{\rm SF}$ if the
Salpeter IMF (0.1--100 $M_\sun$) is assumed.
Since about two-thirds of the ionization-recombination processes produce 
Lyman $\alpha$ photons under case B \citep{spi78}, we also obtain
$L_{\rm Ly \alpha}=0.12fL_{\rm SF}$.
Here, we note that only a fraction $f$ of LC photons ionize hydrogen.
Moreover, we introduce a new parameter, $\gamma$, to represent the
intrinsic stellar luminosity fraction of the SF population to all
stellar population; $\gamma=L_{\rm SF}/L_{\rm total}$, 
where $L_{\rm total}=L_{\rm SF}+L_{\rm under}$.
This parameter is an indicator of the star-forming activity.
Therefore, equation (\ref{eq1}) is reduced to 
\begin{equation}
 \frac{L_{\rm IR}}{L_{\rm SF}}
 =0.37-0.25f+0.63\epsilon+\eta\frac{1-\gamma}{\gamma}\,.
 \label{eq2}
\end{equation}

For dusty starbursts ($\gamma \sim 1$), the right hand side of
equation (\ref{eq2}) becomes unity since the dust opacity is enough high 
to absorb all photons emitted by stars (i.e. $f=0$ and $\epsilon=1$).
In principle, K98 IR formula can be applied to only such galaxies.
On the other hand, if we determine all four parameters ($f$, $\epsilon$,
$\eta$, and $\gamma$), we can convert the observed IR luminosity of all
kinds of galaxies to the SF luminosity, and then, to the current SFR.
This is the algorithm of IHK00.

\section{Dust Opacity of Galaxies}

Here, we describe a model of dust extinction to determine the values of
$f$, $\epsilon$, and $\eta$, in equation (\ref{eq2}).
We will discuss in \S 4.2 the effect of the choice of a specific
extinction law on our result.

Following \cite{cha00} (especially their Fig.\ 1), we consider, first,
two types of dust extinction.
One is the extinction by diffuse ISM dust ($\tau^{\rm ISM}_\lambda$),
and the other is by dust accompanying the birth clouds of young stars
($\tau^{\rm BC}_\lambda$).
Then, we assume 
$\tau^{\rm ISM}_\lambda=\tau^{\rm ISM}_V (\lambda/\lambda_V)^{-n}$ and
$\tau^{\rm BC}_\lambda=\mu\tau^{\rm ISM}_V (\lambda/\lambda_V)^{-n}$ for 
$\lambda > 912$ \AA, where $\mu=2$, $n=0.7$, and $\lambda_V=5500$ \AA\
are adopted, so as to reproduce the observed relation between UV flux
slope and luminosity ratio of IR to UV for nearby star-forming galaxies
\citep{cha00}.
We define the dust optical depth for the SF population as $\tau^{\rm
SF}_\lambda = \tau^{\rm BC}_\lambda + \tau^{\rm ISM}_\lambda$.

For simplicity, we assume that the dust geometry is an uniform screen
(or the optical depth defined above is an effective one).
Then, we define $\epsilon$ as an average of $\epsilon (t)$ over the age
range of 0 -- 10 Myr, where $\epsilon (t)=1-\int L_\lambda (t) \exp
(-\tau_\lambda^{\rm SF}) d\lambda/\int L_\lambda (t) d\lambda$ and
$L_\lambda (t)$ is the luminosity density of a simple stellar population 
(SSP) with an age $t$.
We use SSPs produced by P{\' E}GASE 2.0 \citep{fio97} throughout this
Letter.
In the same way, we define $\eta$ as an average of $\eta(t)$ over the
age of 10 Myr -- 15 Gyr but $\tau_\lambda^{\rm SF}$ is replaced with 
$\tau_\lambda^{\rm ISM}$.
The obtained $\epsilon$ and $\eta$ are a function of only $\tau^{\rm
ISM}_V$, the normalization of the adopted extinction law.

Let us introduce one more type of dust extinction to determine the
parameter $f$, Lyman continuum extinction (Inoue, Hirashita, \&
Kamaya 2001, Inoue 2001), which is the extinction for LC photons by dust
in H {\sc ii} regions before LC photons ionize neutral hydrogens.
The parameter $f$ is determined from the dust optical depth for LC
photons in H {\sc ii} regions, $\tau_{LC}^{\rm HII}$, via eq.\ (8) in
\cite{pet72}.
Here, we assume a scaling law between $\tau_{LC}^{\rm HII}$ and
$\tau^{\rm ISM}_V$, i.e., $\tau_{LC}^{\rm HII}/\tau^{\rm ISM}_V \equiv
\xi$.

Now, we attempt to estimate $\xi$ from real data.
We have shown that $\tau_{LC}^{\rm HII}$ is proportional to the
dust-to-gas mass ratio, $\cal D$, in \cite{ihk01}.
Here, we adopt the relation, $\tau_{LC}^{\rm HII} \simeq ({\cal D}/{\cal
D}_{\rm MW})$ \citep{ino01}, where ${\cal D}_{\rm MW}=6\times 10^{-3}$ is 
a typical Galactic value \citep{spi78}.
We estimated $\cal D$ for seven nearby spiral galaxies observed by ISO
\citep{alt98} in \cite{ihk01}.
The mean value is $1.8 \times 10^{-3}$.\footnote{The mean value is
different from that in \cite{ihk01} because we correct it for He
contribution and for the calibration error of the flux
density at 200 \micron\ (-30\%; \citealt{alt98}). In addition, we
should note that ${\cal D}_{\rm MW}$ by \cite{spi78} is estimated from
the analysis of the extinction properties for a number of line of
sights, which is different from our method.}
On the other hand, we can estimate the dust optical depth from the IR
surface brightness.
By neglecting the self-absorption of the IR photons and assuming the
obtained optical depth to correspond to that in the diffuse ISM, we
obtain $\tau^{\rm ISM}_V =0.36$ as a mean dust optical depth for these
seven galaxies, where we adopt the dust emissivity determined by
\cite{bia99}.
Therefore, we obtain $\xi \approx 1$.

Although the real relation between $\tau_{LC}^{\rm HII}$ and $\tau^{\rm
ISM}_V$ is quite uncertain, we find in equation (\ref{eq2}) that the
effect of $f$ on the ratio $L_{\rm IR}/L_{\rm SF}$ is small.
Indeed, the effect of the deviation from $f=0.5$ on the $L_{\rm
IR}/L_{\rm SF}$ is at most $\pm 0.125$.

\section{Luminosity Ratio of Dust Infrared Emission to Star-Forming
 Population}

Now, we can determine $L_{\rm IR}/L_{\rm SF}$ from
only two parameters; a mean dust opacity ($\tau_V^{\rm ISM}$) and a
star-forming activity ($\gamma$).
We survey these parameters space to clarify when we can determine the
SFR from $L_{\rm IR}$ within a factor of 2 uncertainty via K98 IR
formula, i.e., when $0.5 \le L_{\rm IR}/L_{\rm SF} \le 2.0$ is
satisfied.

In Figure 1, we show some constant $L_{\rm IR}/L_{\rm SF}$ lines in
the $\tau_V^{\rm ISM}$-$\gamma$ plane.
We define the 'applicable area' as the parameter area satisfying the
condition $0.5 \le L_{\rm IR}/L_{\rm SF} \le 2.0$.
The 'applicable area' is between two thick lines in Figure 1 since the
upper thick line is the case of $L_{\rm IR}/L_{\rm SF}=0.5$ and the
lower one is the case of $L_{\rm IR}/L_{\rm SF}=2.0$.
Also, the thin line in Figure 1 represents the case of $L_{\rm
IR}/L_{\rm SF}=1.0$.
We find that the 'applicable area' is very wide in the $\tau_V^{\rm
ISM}$-$\gamma$ plane.
This is a main conclusion in this Letter.

For a galaxy with an exponentially declining star formation history
(SFH) of $\ga$ 10 Gyr time scale, age of $\ga$ 1 Gyr, and Salpeter IMF,
we find that $\gamma \sim 0.5$.
Thus, many spiral and irregular galaxies are likely to fall in the
'applicable area'.
In order to examine whether real galaxies fall in the 'applicable area'
or not, we will estimate their $\gamma$ and $\tau_V^{\rm ISM}$ in the
following.
Then, we will discuss implications from our results.

\subsection{Estimation of $\gamma$ and $\tau_V^{\rm ISM}$}

First, we estimate $\gamma$ of galaxies from the observed equivalent
width of H $\alpha$ emission line corrected for dust and stellar
absorption effects.
By the definition of $\gamma$, we can express $\gamma = EW_{\rm H\alpha} 
(C_{\rm H\alpha}/C_R)$, where $EW_{\rm H\alpha}$ is the intrinsic
equivalent width of H $\alpha$ emission line, $C_{\rm H\alpha}$ is the
intrinsic ratio of the bolometric luminosity to the H $\alpha$ emission
line luminosity of the SF population, and $C_R$ is the intrinsic ratio of
the bolometric luminosity to the luminosity density at $R$-band of all
stellar population.

If we specify the metallicity and IMF of galaxies, the value of $C_{\rm
H\alpha}$ is determined from only the electron temperature and
density in H {\sc ii} regions under case B approximation.
On the contrary, the value of $C_R$ depends on the SFH and age of
galaxies as well as the metallicity and IMF.
We examine the evolution of $C_R$ with the galactic age, so that we find
that the factor $C_{\rm H\alpha}/C_R$ falls in the range of 0.002 --
0.004 unless we consider a galaxy younger than a few Gyr or a galaxy
with a short ($\la$ 1 Gyr) exponential time scale.
Thus, we adopt a constant 0.003 as $C_{\rm H\alpha}/C_R$, which
corresponds to the case of Solar metallicity, exponentially declining
SFH with 10 Gyr time scale, and about 13 Gyr age.
Here, we do not take into account the chemical enrichment effect.

Therefore, we estimate $\gamma$ from the following equation
approximately;
\begin{equation}
 \gamma = 
  \cases{0.003 EW_{\rm H\alpha} & for $EW_{\rm H\alpha} \le 333$ \AA\cr
         1 & for $EW_{\rm H\alpha} > 333$ \AA}\,,
 \label{eq3}
\end{equation}
where the second case is needed due to $\gamma \le 1$ by its definition.
In the real case, the value of $\gamma$ asymptotically approaches unity
by decreasing the factor $C_{\rm H\alpha}/C_R$, when $EW_{\rm H\alpha}$
increases.
Thus, we overestimate $\gamma$ for a galaxy with a very large
$EW_{\rm H\alpha}$.
However, the real value of $\gamma$ is about unity for such galaxies.
On the other hand, we underestimate $\gamma$ for a galaxy with a very
small $EW_{\rm H\alpha}$.
We confirm that this effect is also not so large (at most factor of 2).

Although the effect of the change of the metallicity on $C_{\rm
H\alpha}/C_R$ is small, that of the IMF slope is large.
If we choose the IMF slope of 1.35 (Salpeter's slope is 2.35), a typical 
$C_{\rm H\alpha}/C_R$ decreases an order of magnitude, about 0.0003.
In the case of the slope of 3.35, we find $C_{\rm H\alpha}/C_R \simeq
0.007$ typically.
This is because a top-heavy IMF (1.35 slope) results in a smaller
fraction of the $R$-band luminosity density in the bolometric
luminosity since massive stars do not contribute to the $R$-band light
so effectively.
A steep IMF provides an opposite result.

Now, we must estimate the intrinsic equivalent width, $EW_{\rm
H\alpha}$, from the observed one, $EW_{\rm H\alpha}^{\rm obs}$.
Assuming that the underlying stellar population (age $\ga$ 10 Myr) 
dominates the $R$-band continuum and the stellar Balmer absorption, we
can express 
\begin{equation}
 EW_{\rm H\alpha}=(EW_{\rm H\alpha}^{\rm obs}+EW_{\rm H\alpha}^{\rm *,abs})
                  \frac{1}{f}
                  \exp (\tau^{\rm SF}_{\lambda_{\rm H\alpha}}
                        - \tau^{\rm ISM}_{\lambda_{\rm H\alpha}})\,,
 \label{eq4}
\end{equation}
where $EW_{\rm H\alpha}^{\rm *,abs}$ is the equivalent width of the
stellar absorption, and $\lambda_{\rm H\alpha}=6563$ \AA.
The factor $1/f$ means that only a fraction $f$ of LC photons is used
to ionize neutral hydrogen atoms.

Next, we estimate $\tau_V^{\rm SF}$ from the Balmer decrement corrected
for the stellar absorption by assuming $EW_{\rm H\alpha}^{\rm
*,abs}=EW_{\rm H\beta}^{\rm *,abs}=5$ \AA, which noted by \cite{ken92}
as a typical value, and then, we convert the obtained $\tau_V^{\rm SF}$
into $\tau_V^{\rm ISM}$ by using the parameter $\mu=2$ (See \S 3).
Here, we also assume the electron temperature of $10^4$ K and case B.
Once the optical depth $\tau_V^{\rm ISM}$ is obtained, we can estimate
$f$ parameter, $EW_{\rm H\alpha}$ from $EW_{\rm H\alpha}^{\rm obs}$, and 
then, $\gamma$.
The obtained values of $\tau_V^{\rm ISM}$ are systematically smaller
than  but roughly consistent with those determined from the UV slope
method \citep{meu99} for sample galaxies described below.

The uncertainty of $\tau_V^{\rm ISM}$ from the Balmer decrement 
is large, $\sim \pm 0.2$.
As a result, the uncertainty of the obtained $\gamma$ is $\sim \pm 0.2$, 
which is estimated from a simple Monte Carlo simulation if we do not
take account of the uncertainties of the adopted parameters of the dust
extinction law (i.e., $\eta$, $\xi$, and $n$ in \S 3).
Also, we find that a random error of $\tau_V^{\rm ISM}$ causes a
somewhat overestimation of $\gamma$.
To display this point, we show a typical error bar in Figure 1.
In addition, we select only galaxies with enough strong H $\beta$
'emission' line to obtain a reliable Balmer decrement.
Thus, our sample galaxies are biased towards larger $\gamma$.

One might think that the parameter $\gamma$ can be estimated from the
ratio of the UV luminosity and the $R$ or $K$-band luminosity.
However, we can do it reasonably only if we specify the SFH and age of
galaxies.
That is, the coefficient between $\gamma$ and $L_{UV}/L_R$ or
$L_{UV}/L_K$ varies together with the galactic age significantly.
This is because a significant fraction of the UV light originates in the 
underlying stellar population with its age of $\ga$ 10 Myr.
On the contrary, the ionizing photons producing the H $\alpha$ photons
originate in almost only the SF population (its age $\la$ 10 Myr).
Therefore, we estimated $\gamma$ from the equivalent width of the H
$\alpha$ line.

\subsection{Results \& Discussions}

To plot real galaxies in Figure 1, we compile all morphological types of
galaxies except for elliptical galaxies from the literature.
In Figure 1, the filled circles are early type spiral (Sa--Sab) galaxies
observed by \cite{usu01}, who selected active star-forming early type
spiral galaxies with $\log(L_{\rm FIR}/L_B) \ge 0.5$.
The filled squares, open squares, and crosses are early type spiral
(S0--Sbc), late type spiral (Sc--Sm), and irregular (including peculiar
and merger) galaxies, respectively, observed by \cite{ken92}, who
selected galaxies covering all morphological types.
The filled triangles, open triangles, and stars are also early spiral,
late spiral, and irregular (also peculiar and merger) galaxies,
respectively, observed by \cite{sto95}, who selected active star-forming 
galaxies observed by $IUE$ satellite.
Any galaxies containing an AGN are excluded.

We find that almost all (43/52) galaxies fall in the 'applicable area'.
For 7 out of 9 galaxies in the outside of the 'applicable area', we fail 
to determine $\tau_V^{\rm ISM}$ because they become negative values.
Although the uncertainties of the determined $\tau_V^{\rm ISM}$ and
$\gamma$ are large and compiled galaxies do not make a complete sample,
this result indicate that many nearby normal galaxies are really in the
'applicable area'.

We emphasize the following two points.
First, it is a just coincidence that K98 IR formula is applicable to
normal galaxies.
It results from offsetting two effects of a small dust opacity and a
large cirrus contribution of normal galaxies relative to starburst
galaxies on the conversion coefficient of K98 IR formula (Kennicutt
1998b, IHK00).
For an example case of a galaxy with $\tau_V^{\rm ISM}=0.3$ and $EW_{\rm 
H\alpha}^{\rm obs}= 50$ \AA, while only about 70\% luminosity of the SF
population is converted into the IR luminosity, it is almost the same
amount as the rest 30\% SF luminosity that is supplied with the IR
luminosity by the underlying population.

Second, the result above is robust for changing the adopted parameter
sets of the dust extinction.
One might think that our result depends on the dust extinction
law adopted.
However, we note that it dose not change when we change the parameters
in the range of $0 \le \mu \le 3$, $0.4 \le n \le 1.0$, and $0 \le
\xi \le 2$ as shown in Figure 2 (an extra figure).
This point will be discussed in detail in the forthcoming paper
(Inoue 2002, in preparation).\footnote{Visit the following web cite
where you will find figures showing the cases of other parameter choice; 
http://www.kusastro.kyoto-u.ac.jp/\~{}inoue/irsfr.html.}

We discuss the condition required to fall in the 'applicable area'.
{}From the lower bound of $\gamma$ in Figure 1 ($L_{\rm IR}/L_{\rm
SF}=2.0$), we find that a quiescent galaxy with a large dust amount (a
small $\gamma$ and a large $\tau_V^{\rm ISM}$) is in the outside of the
'applicable area'.
Since galaxies showing a lower activity of the star formation tend to
be more transparent \citep{wan96,hop01}, the number of galaxies falling
in the area of large $\tau_V^{\rm ISM}$ and small $\gamma$ may be small.
Moreover, we find that the lower bound of $\gamma$ is $\sim 0.1$ for a
galaxy with $\tau_V^{\rm ISM} \sim 0.3$.
This corresponds to $EW_{\rm H\alpha}^{\rm obs} \ga 10$ \AA.
This boundary can be exceeded not only by active star-forming galaxies
but also by many nearby normal spiral and irregular galaxies
\citep{ken83}.

The upper bound of $\gamma$ in Figure 1 ($L_{\rm IR}/L_{\rm SF}=0.5$)
reaches unity if $\tau_V^{\rm ISM}$ is larger than about 0.1.
Thus, a starburst galaxy having very small dust content of
$\tau_V^{\rm ISM} \la 0.1$ is in the outside of the 'applicable area'.
For example, I Zw 18, which is an active star-forming and the most metal
deficient nearby dwarf galaxies, has typically $A_V=0.2$ mag in the
southeast H {\sc ii} region \citep{can02}.
It corresponds to $\tau_V^{\rm ISM} \sim 0.06$ if we adopt $\mu=2$.
By taking into account $\gamma \sim 1$ estimated from very large
$EW_{\rm H \alpha}^{\rm obs} \sim 1000$ \AA\ \citep{can02}, I Zw 18 is
in the outside of the 'applicable area'.

In conclusion, we find that the range of a dust opacity and a
star-forming activity to be $0.5 \le L_{\rm IR}/L_{\rm SF} \le 2.0$ is
very large.
Furthermore, we confirm that these parameters of many nearby normal and
active star-forming galaxies really fall in this range.
Therefore, we can apply K98 IR formula assuming $L_{\rm IR}=L_{\rm
SF}$ to many nearby normal galaxies as well as dusty starbursts 
without any correction factor within a factor of 2 uncertainty.
However, we cannot apply it to dust rich but quiescent galaxies and
extremely dust poor galaxies.
Therefore, we need to divide the coefficient of K98 IR formula by a
correction factor, $L_{\rm IR}/L_{\rm SF}$, when we apply it to such
galaxies.
Following the model in this Letter, we can estimate the correction
factor for any galaxies if we have all data required, although we must
more refine the model.

\acknowledgments

The author would like to thank H. Kamaya, T. T. Takeuchi, and
H. Hirashita for their suggestions to improve the quality of this work, 
and also, to thank an anonymous referee for his/her many helpful
comments to this paper.
The author has made extensive use of NASA's Astrophysics Data System
Abstract Service (ADS).

\begin{figure}
\plotone{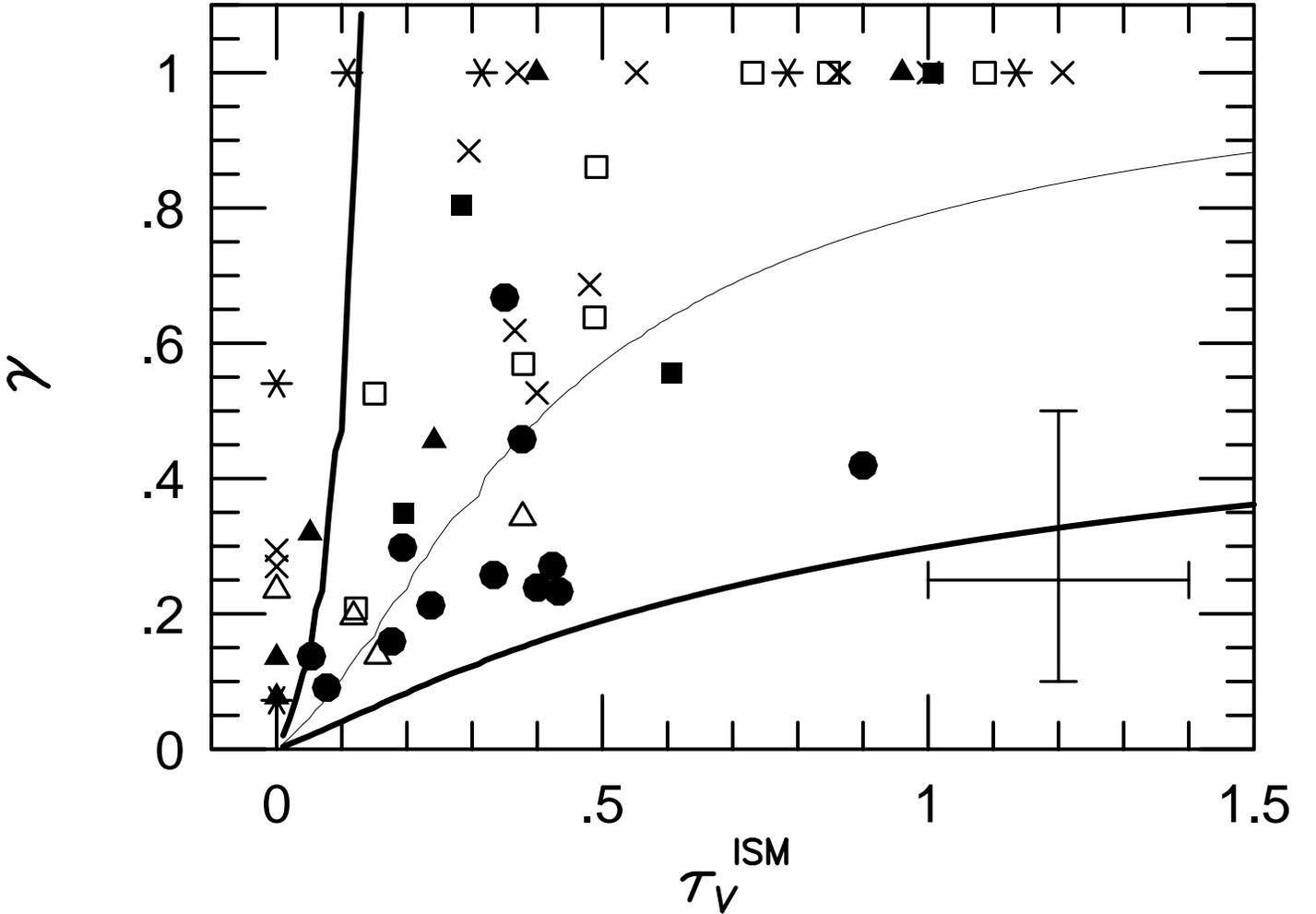}
\figcaption{Iso-$L_{\rm IR}/L_{\rm SF}$ lines in the $\tau_V^{\rm 
 ISM}$-$\gamma$ plane, where $L_{\rm IR}$ is the observed total IR
 luminosity, $L_{\rm SF}$ is the intrinsic bolometric luminosity of
 newly ($\la$ 10 Myr) formed stellar population, $\tau_V^{\rm ISM}$ is a 
 mean optical depth by dust in the diffuse interstellar medium, and
 $\gamma$ is the luminosity fraction of newly formed stellar
 population. The upper thick line is $L_{\rm IR}/L_{\rm SF}=0.5$, and
 the lower thick line is $L_{\rm IR}/L_{\rm SF}=2.0$. The thin line
 is $L_{\rm IR}/L_{\rm SF}=1.0$. The filled circles are early type
 spiral (Sa--Sab) galaxies taken from \cite{usu01}. The filled squares,
 open squares, and crosses are early type spiral (S0--Sbc), late type
 spiral (Sc--Sm), and irregular (including peculiar and merger)
 galaxies, respectively, taken from \cite{ken92}. The filled triangles,
 open triangles, and stars are also early, late spirals, and irregular
 galaxies, respectively, taken from \cite{sto95}. The classifications
 are taken from NASA/IPAC Extragalactic Database. No galaxies with an
 AGN are included. A typical error bar is shown in the bottom right of
 the panel.}
\end{figure}

\begin{figure}
\plotone{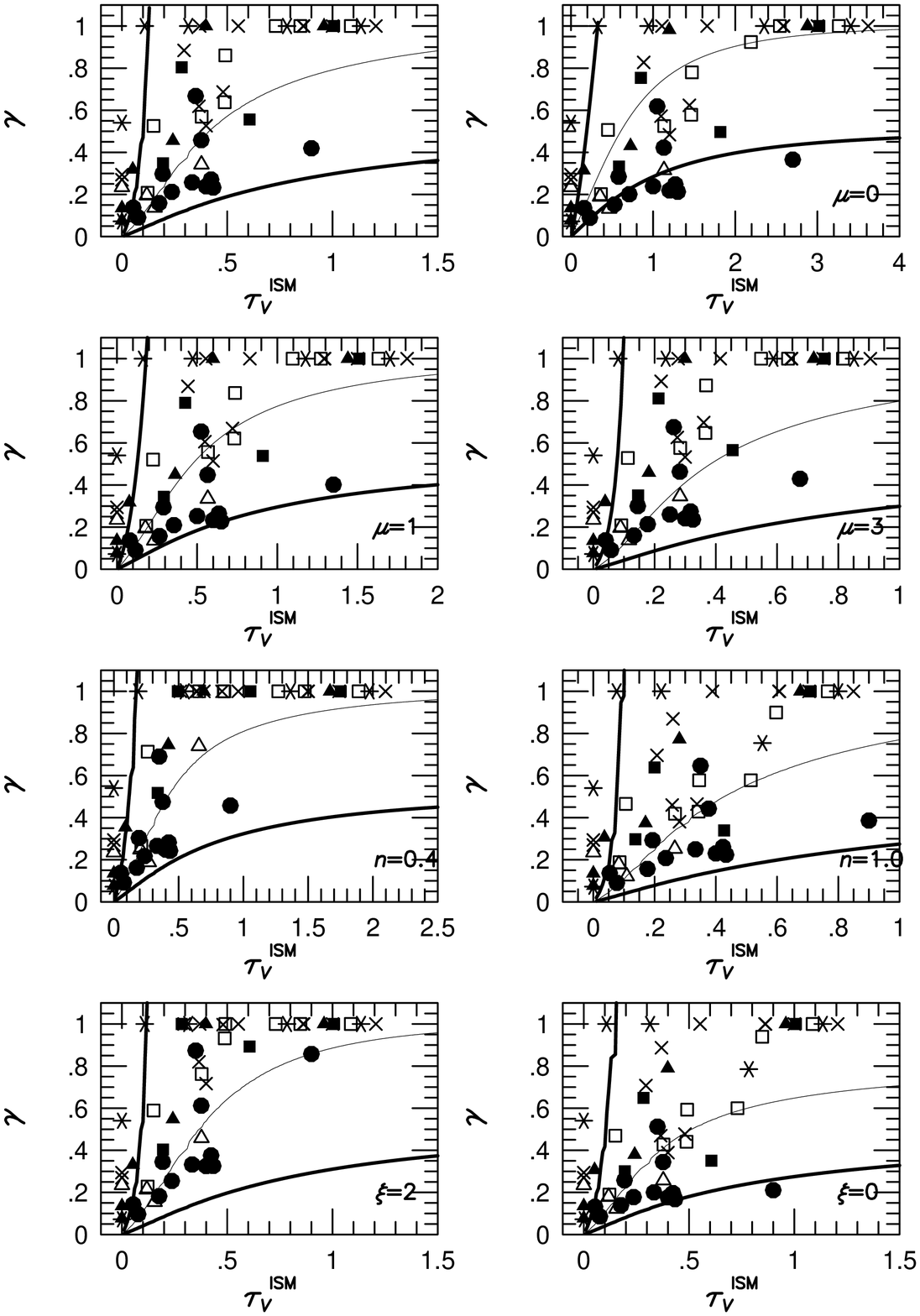}
\figcaption{Same as Fig 1, but the cases adopted other parameters choice 
 of the extinction law.}
\end{figure}


\begin{thebibliography}{}
\bibitem[Alton et al.(1998)]{alt98}
 Alton, P. B., Trewhella, M., Davies, J. J., Evans, R., Bianchi, S.,
	Gear, W., Thronson, H., Valentijn, E., \& Witt, A.  1998, \aap,
	335, 807

\bibitem[Bell \& Kennicutt(2001)]{bel01}
 Bell, E., \& Kennicutt, R. C.  2001, \apj, 548, 681

\bibitem[Bianchi, Davies, \& Alton(1999)]{bia99}
 Bianchi, S., Davies, J. I., \& Alton, P. B.

\bibitem[Buat \& Xu(1996)]{bua96}
 Buat, V., \& Xu, C.  1996, \aap, 306, 61

\bibitem[Buat et al.(2002)]{bua02}
 Buat, V., Boselli, A., Gavazzi, G., \& Bonfanti, C.  2002, \aap, 383,
	801

\bibitem[Cannon et al.(2002)]{can02}
 Cannon, J. M., Skillman, E. D., Garnett, D. R., \& Dufour, R. J.  2002, 
	\apj, 565, 931

\bibitem[Charlot \& Fall(2000)]{cha00}
 Charlot, S., \& Fall, S. M.  2000, \apj, 539, 718 

\bibitem[Charlot et al.(2002)]{cha02}
 Charlot, S., Kauffmann, G., Longhetti, M., Tresse, L., White, S. D. M., 
	Maddox, S. J., \& Fall, S. M.  2002, \mnras, 330, 876

\bibitem[Fioc \& Rocca-Volmerange(1997)]{fio97}
 Fioc, M., \& Rocca-Volmerange, B.  1997, \aap, 326, 950

\bibitem[Hopkins et al.(2001)]{hop01}
 Hopkins, A. M., Connolly, A. J., Haarsma, D. B., \& Cram, L. E.  2001,
	\aj, 122, 288

\bibitem[Inoue, Hirashita, \& Kamaya(2000)]{ihk00}
 Inoue, A. K., Hirashita, H., \& Kamaya, H.  2000, \pasj, 52, 539 (IHK00)

\bibitem[Inoue et al.(2001)]{ihk01}
 Inoue, A. K., Hirashita, H., \& Kamaya, H.  2001, \apj, 555, 613

\bibitem[Inoue(2001)]{ino01}
 Inoue, A. K.  2001, \aj, 122, 1788

\bibitem[Kennicutt(1983)]{ken83}
 Kennicutt, R. C.  1983, \apj, 272, 54

\bibitem[Kennicutt(1992)]{ken92}
 Kennicutt, R. C.  1992, \apj, 388, 310

\bibitem[Kennicutt(1998a)]{ken98a}
 Kennicutt, R. C.  1998a, \apj, 498, 541 (K98)

\bibitem[Kennicutt(1998b)]{ken98b}
 Kennicutt, R. C.  1998b, \araa, 36, 189

\bibitem[Meurer, Heckman, \& Calzetti(1999)]{meu99}
 Meurer, G. R., Heckman, T. M., \& Calzetti, D.  1999, \apj, 521, 64

\bibitem[Petrosian, Silk, \& Field(1972)]{pet72}
 Petrosian, V., Silk, J., \& Field, G. B. 1972, \apj, 177, L69

\bibitem[Rosa-Gonz{\'a}lez et al.(2002)]{ros02}
 Rosa-Gonz{\'a}lez, D., Terlevich, E., \& Terlevich, R.  2002, \mnras,
	in press (astro-ph/0112556)

\bibitem[Sanders \& Mirabel(1996)]{san96}
 Sanders, D. B., \& Mirabel, I. F.  1996, \araa, 34, 749

\bibitem[Soifer, Houck, \& Neugebauer(1987)]{soi87}
 Soifer, B. T., Houck, J. R., \& Neugebauer, G.  1987, \araa, 25, 187

\bibitem[Spitzer(1978)]{spi78}
 Spitzer, L.  1978, Physical Processes in the Interstellar Medium
	(New York: Wiley)

\bibitem[Storchi-Bergmann, Kinney, \& Challis(1995)]{sto95}
 Storchi-Bergmann, T., Kinney, A. L., \& Challis, P.  1995, \apjs, 98, 103

\bibitem[Sullivan et al.(2000)]{sul00}
 Sullivan, M., Treyer, M. A., Ellis, R. S., Bridges, T. J., Milliard,
	B., \& Donas, J.  2000, \mnras, 312, 442

\bibitem[Usui, Sait{\= o}, \& Tomita(2001)]{usu01}
 Usui, T., Sait{\= o}, M., \& Tomita, A.  2001, \aj, 121, 2483

\bibitem[Wang \& Heckman(1996)]{wan96}
 Wang, B., \& Heckman, T. M.  1996, \apj, 457, 645

\end{thebibliography}
\end{document}